\documentstyle[12pt]{article}
\begin{document}
\begin{titlepage}
\begin{flushright}
\begin{tabular}{l}
IFT/8/95\\
hep-ph/9506462\\
June 1995
\end{tabular}
\end{flushright}
\vspace{2.0cm}
\begin{center}
{\large
IMPROVED EVALUATION OF THE NNLO QCD EFFECTS IN THE TAU DECAY,
$e^{+}e^{-}$ ANNIHILATION INTO HADRONS AND DEEP-INELASTIC
SUM RULES\\
}
\vspace{1.3cm}
 Piotr A. R\c{a}czka\\
 Institute of Theoretical Physics,  Department of Physics\\
 Warsaw University,  ul.\ Ho\.{z}a 69, 00-681 Warsaw, Poland\\
 E-mail: praczka@fuw.edu.pl\\
\vspace{2.5cm}
{\bf Abstract\\[5pt]}
\parbox[t]{\textwidth}{\small
A systematic method is proposed for analyzing the
renormalization scheme uncertainties
in the next-next-to-leading order QCD predicitions,
based on a condition which eliminates schemes that give
rise to large cancellations in the expression for the
characteristic scheme invariant combination of the
expansion coefficients.
Using this method it is shown that the QCD corrections to
the tau lepton decay are rather stable with respect to
change of the scheme, provided that an improved formula
is used, which involves numerical evaluation of the
contour integral in the complex energy plane with the
Adler function under the integral. Optimized predictions
for the tau decay corrections are given. It is shown that
also in the case of the of QCD corrections to $e^{+}e^{-}$ annihilation into
hadrons the conventional expansion has sizable
scheme dependence, even at large
energies. However, a considerable improvement is obtained
when the QCD corrections are expressed as a contour integral,
 with the Adler function under the integral, resumming in
this way the large $\pi^{2}$ contributions.
In the case of the corrections to the Bjorken sum rule
for polarized structure functions it is found that for $n_{f}=4$
they are insensitive to change of the scheme.
However, the $n_{f}=3$ expression
is found to be strongly scheme dependent at lower energies.
}
\vfill
\end{center}
{\small Talk presented at the XXXth Rencontres de Moriond
``QCD and High Energy Hadronic Interactions,''
Les Arcs, France, March 19-26, 1995,
to appear in the proceedings.}
\end{titlepage}
\newpage
\bigskip

 Perturbative predictions in finite order of
 perturbation expansion depend on the choice of the renormalization
 scheme (RS).  In the
 next-to-next-to-leading order (NNLO) approximation, when mass effects are
 neglected, the freedom of choice of RS  may be characterized
 by two independent continuous parameters. The differences
 between predictions
 in various schemes are formally of higher order in the coupling
 constant, but numerically they are significant for phenomenology.

 There are two things that one should do with the
  RS dependence of perturbative approximants.
 First, using various heuristic arguments of
 physical or technical character,  one should make a careful choice of
 the scheme, obtaining an ``optimized'' perturbative prediction.
 Several propositions have been discussed in the literature
 \cite{DukeRoberts}, including a very interesting prescription based on the
 Principle of Minimal Sensitivity (PMS) \cite{PMS}. Secondly,
 one should investigate how strongly predictions change when one
 moves away from the preferred RS. By calculating the variation
 in the predictions over a set of {\em a priori} acceptable schemes
 one obtains  an estimate of the  reliability of the
 ``optimized'' prediction. A systematic method for obtaining such
 estimates has been recently presented in \cite{par95}.
 This method is based on
 a condition which eliminates from the analysis the schemes
 that give rise to unnaturally large  expansion coefficients
 in the expansion for the
 physical quantity or the beta-function.

 Let us consider a NNLO expression for a physical quantity $\delta$,
 depending on one energy variable $P$, in the massless approximation:
\begin{equation}
\delta(P^{2}) = a(P^{2})[1+r_{1}a(P^{2})+r_{2}a^{2}(P^{2})],
\label{delta}
\end{equation}
 where $a(\mu^{2})=g^{2}(\mu^{2})/(4 \pi^{2})$
 denotes the coupling constant that satisfies the NNLO
 renormalization group equation:
\begin{equation}
\mu \frac{da}{d\mu} = - b\,a^{2}\,(1 + c_{1}a + c_{2}a^{2}\,),
\label{rge}
\end{equation}
 The expansion coefficients $r_{i}$ and $c_{2}$ depend on the choice
 of the renormalization scheme --- the relevant formulas have been
 collected for example in \cite{par92} --- but there exists a combination
 of these coefficients which is independent of the
 scheme:
\begin{equation}
\rho_{2}=c_{2}+r_{2}-c_{1}r_{1}-r_{1}^{2}.
\label{rho2}
\end{equation}
 As was discussed in \cite{par92,par95}, this combination provides a
 natural RS independent
 characterization of the magnitude of the NNLO correction.
 The RS invariant $\rho_{2}$ may be used
 to distinguish between  ``good'' and ``bad'' schemes.  Indeed,
 one usually identifies unnatural schemes with ones that introduce
 large expansion coefficients in an artificial way. However, the
 combination (\ref{rho2}) must stay the same even for very bad schemes,
 which may be achieved only by the presence of large cancellations
 between various terms in the expression for $\rho_{2}$ \cite{par92}.
 It is then useful to introduce a function:
\begin{equation}
\sigma_{2}(r_{1},r_{2},c_{2})=|c_{2}|+|r_{2}|+c_{1}|r_{1}|+r_{1}^{2}.
\end{equation}
 This function  measures the degree of cancellation in $\rho_{2}$,
 and gives a quantitative meaning to the notion of
``naturalness'' of any chosen RS.
 It is clear that a large cancellation in $\rho_{2}$ would
 imply that the value of $\sigma_{2}$ would be large compared to
 $|\rho_{2}|$.
 (Obviously $\sigma_{2}\geq|\rho_{2}|$.) If we  have any preference
for using some renormalization  scheme, we should also include in the
analysis predictions obtained in schemes which have the same, or
smaller,  value of $\sigma_{2}$.
For example, for the PMS scheme we have:
\begin{equation}
r_{1}^{PMS}=O(a^{PMS}),\,\,\,\,
c_{2}^{PMS}=1.5\rho_{2}+O(a^{PMS}),
\end{equation}
which implies $\sigma_{2}(PMS)\approx 2|\rho_{2}|$.
More generally, we may write the condition
on the acceptable schemes in the form:
\begin{equation}
\sigma_{2}(r_{1},r_{2},c_{2}) \leq l\,|\rho_{2}|.
\label{constraint}
\end{equation}
 The constant $l\,$ in the
 condition (\ref{constraint}) controls the degree of cancellations
 that we want to allow in the expression for $\rho_{2}$. The value
 $l=2$ in (\ref{constraint})
 is the minimal value of $l\,$ for which the PMS scheme
 falls into the ``allowed''  region of scheme parameters.
 The proposition of \cite{par92,par95} is to calculate the variation
 of the predictions for $\delta$ over the set of
schemes satisfying the condition
 (\ref{constraint}), and use this variation as a quantitative
 estimate of reliability of the perturbative predictions.
 As was discussed in \cite{par92}, it is convenient to use
 $r_{1}$ and $c_{2}$ to parametrize the freedom of choice of the
 RS in the NNLO approximants.

It should be stressed that no claim is made that in this way
the actual theoretical error of the calculation is obtained, i.\ e.\
there is of course no theorem that guarantees that
the difference between the NNLO prediction and the true
result would lie within the obtained ``error bound.''
However, the region of scheme parameters satisfying
the condition (\ref{constraint}) with $l=2$ appears to be the
minimal set of schemes that has to be taken into account ---
consequently, if strong scheme dependence is obtained for these
schemes, it is an unambiguos sign that the perturbation series
is not reliable. Also, evaluating by our method the
variation in predictions
for several quantities with the same value of $l$ in the
condition (\ref{constraint}) we obtain a very good estimate of
{\em relative} reliability of the predictions. The obtained
estimates may then be used for example to assign different
weights to different observables in the global fit of
$\Lambda_{\overline{MS}}$.

Let us begin with the QCD correction
 $\delta_{\tau}$  to the tau lepton decay into hadrons.
 The QCD prediction for  $\delta_{\tau}$
 may be expressed as a contour integral in the complex
 energy plane \cite{bra89}:
\begin{equation}
\delta_{\tau}=\frac{1}{2\pi}\int_{-\pi}^{\pi}d\theta
\left(1+2e^{i\theta}-2e^{3i\theta}-e^{4i\theta}\right)
\left[\delta_{\Pi}(-\sigma)|_{\sigma=-m_{\tau}^{2}e^{i\theta}}\right],
\label{rtcont}
\end{equation}
where $\delta_{\Pi}(-\sigma)$ is the so called Adler function, which
 has the form (\ref{delta}). The contour integral has been initially
 evaluated by expanding $a(-\sigma)$ in terms of
 $a(m_{\tau}^{2})$.
 This led to the frequently used expansion for
 $\delta_{\tau}$ in terms of $a(m_{\tau}^{2})$, which
 has the form (\ref{delta}), with $\rho_{2}^{\tau}=-5.476$
\cite{kataev,samuel}.
  The region of scheme
parameters satisfying the condition (\ref{constraint}) with $l=2$ is
approximately given by $r_{1}\in (-1.54,2.11)$ and
 $c_{2}\in (-8.21,2.74)$.
 Applying  to this expansion  the
 method  outlined above we find very strong RS dependence.
 This confirms observations made in  \cite{par92}.
 However, as was pointed out
 in \cite{piv,ledepi}, one may evaluate $\delta_{\tau}$ in an improved way,
 using under the contour integral the renormalization group
 improved expression for $\delta_{\Pi}$, and calculating the contour
 integral numerically. In this way one resumms large
 $\pi^{2}$ corrections which would otherwise appear in higher orders.
 Let us note that for $\delta_{\Pi}$ one has $\rho_{2}^{\Pi}=5.238$, which
 has the same magnitude, but an opposite sign compared to value
 obtained for the  ``naive'' expansion.
 A complete analysis of the RS dependence of the improved
 expression for $\delta_{\tau}$ has been described in \cite{parszym94}.
 (\cite{ledepi}  contains some discussion of scale
 dependence of the improved expression.)
 The QCD predictions obtained in the improved evaluation
 appear to be quite stable with respect to change of RS, despite
 the low energy scale of the process. The problem of finding
 the  PMS predictions for the improved expression has been
 considered. This is nontrivial since for the improved approximant
 we cannot use the set of algebraic PMS equations given in \cite{PMS}.
 It was found that the
  location of the critical point
 closest to the $l=2$ region of allowed scheme parameters is
 well approximated by $r_{1}=0$ and  $c_{2}=1.5\,\rho_{2}^{\Pi}$, for
 most values of  $m_{\tau}/\Lambda_{\overline{MS}}$.
 Therefore predictions in this scheme have been taken as preferred
   predictions for the phenomenological analysis.
 Assuming $(R_{\tau})_{exp}=3.591\pm0.036$ as an averaged experimental
 value (see discussion of the experimental results
  in the Appendix of \cite{parszym94}) it was found that
\begin{math}
\Lambda^{(3)}_{\overline{MS}}=376(\mbox{opt})
^{+15}_{-14}(\mbox{th,l=2})\pm29(\mbox{exp})\,\mbox{MeV}.
\end{math}
For the $l=3$ region the variation is $^{+26}_{-21}\,\mbox{MeV}$.
This corresponds to:
\begin{math}
\alpha_{s}^{\overline{MS}}(m_{\tau}^{2})=
0.332^{+0.008}_{-0.007}(\mbox{th,l=2})\pm0.015(\mbox{exp}).
\end{math}
For the $l=3$ region we obtain variation of $^{+0.014}_{-0.010}$.
Extrapolating to $m_{Z}^{2}$ we find
\begin{math}
\alpha_{s}^{\overline{MS}}(m_{Z}^{2})=
0.1190^{+0.0009}_{-0.0008}(\mbox{th,l=2})
\pm0.0017(\mbox{exp}).
\end{math}
(For the $l=3$ region the variation is $^{+0.0016}_{-0.0013}$.)
We see that the RS
dependence ambiguities in the determination of
$\alpha_{s}^{\overline{MS}}(m_{\tau}^{2})$
 or $\alpha_{s}^{\overline{MS}}(m_{Z}^{2})$
 from $R_{\tau}$ are
 not very big --- this is the result of stabilization of
the predictions when improved evaluation procedure is used.
It should be noted however that the RS dependence ambiguities are still
 comparable in magnitude to the
 uncertainties related to  the
 present experimental accuracy of $R_{\tau}$.

In the case of the QCD correction $\delta_{e^{+}e^{-}}$ to the
the $e^{+}e^{-}$ annihilation into
hadrons, for $n_{f}=5$, we have $\rho_{2}^{e^{+}e^{-}}=-15.055$
\cite{kataev, samuel} (this does not include a very small
singlet correction, which is added separately).
 The region of scheme
parameters satisfying the condition (\ref{constraint}) with $l=2$ is
approximately given by $r_{1}\in (-4.8,4.2)$ and
 $c_{2}\in (-22.6,7.5)$, and for $l=3$ it is approximately
 $r_{1}\in (-5.5,4.9)$ and  $c_{2}\in (-30.2,15.1)$. Changing the
scheme parameters in the $l=2$ region we find for
$s/\Lambda^{2}_{\overline{MS}}=75^{2}$ the variation in
the predictions from 0.0491 to 0.0537, and for the $l=3$ region
from 0.0454 to 0.0539. We see that even though the considered energy is
high and the perturbation series should be reliable, the
scheme dependence of the predictions is surprisingly large.
This is a consequence of the fact that $\delta_{e^{+}e^{-}}$
has very large NNLO correction, which is reflected by the
magnitude of the RS invariant. However, closer examination shows
that major part of the NNLO correction comes from the term
proportional to $\pi^{2}$, which appears in the process of
analytic continuation from spacelike to timelike momenta.
The large contributions from the $\pi^{2}$ terms may be avoided
--- similarly as in the case of the tau decay ---
by expressing $\delta_{e^{+}e^{-}}$ as a contour integral in the
complex energy plane, with the renormalization group improved
expression for the Adler function under the integral \cite{radu}:
\begin{equation}
\delta_{e^{+}e^{-}}(s)=\frac{1}{2\pi}\int_{-\pi}^{\pi}d\theta\,
\delta_{\Pi}(-\sigma)|_{\sigma=-s\,e^{i\theta}}.
\end{equation}
 For the
Adler function for $n_{f}=5$ we have $\rho_{2}^{\Pi}=-2.969$.
The improved expression
for $\delta_{e^{+}e^{-}}$ appears to be much more stable with respect to
change of RS as compared to the conventional expression
 \cite{parszym95}. Also, the
$l=2,3$ regions of the scheme parameters are much smaller than in
the case of conventional expansion, because the RS invariant is much
smaller. In Fig.1 we show the contour plot of $\delta_{e^{+}e^{-}}$,
as a function of scheme parameters, for
$s/\Lambda^{2}_{\overline{MS}}=75^{2}$. We see that the variation
of predictions over the $l=2$ region of scheme parameters is
in fact negligible from the point of view
of phenomenological applications. For $n_{f}=3,4$ the same
effect of reduced RS dependence in the improved formula
is found.  Incidentally, for $n_{f}=5$, the optimized
prediction obtained from the contour integral is close to
the prediction obtained with the conventional expansion in the
$\overline {MS}$ scheme and the PMS prediction in the
conventional expansion.
This is not the case for other number of flavors.

In the case of the QCD correction to the Bjorken sum rule for the
polarized structure functions \cite{larin} it was found that the
expression for $n_{f}=4$, with $\rho_{2}=1.330$ is very stable with
respect to the change of the RS. The situation
with $n_{f}=3$ predictions is quite different.
For $n_{f}=3$ we have $\rho_{2}=5.476$, so
that the $l=2$ region of scheme parameters extends approximately
for $r_{1}\in (-1.65,1.0)$ and  $c_{2}\in (-2.8,8.3)$. Changing the scheme
parameters in this region we find a considerable variation of the
predictions for $Q^{2}/\Lambda_{\overline {MS}}^{2}$ below
approximately $6^{2}$, as is shown in Fig.2.
 This indicates that perturbative predictions
in this region are not reliable, even though we may still obtain
PMS predictions. Unfortunately, in
the case of the Bjorken sum rule we cannot expect that the RS
dependence would be reduced in a way similar to tau decay or
$e^{+}e^{-}$ annihilation. Other ways of improving
the QCD perturbation expansion in this case have to be investigated.
It should be noted that the QCD correction to the
Gross-Llewellyn-Smith sum rule has almost identical expansion
as in the case of the Bjorken sum rule, so that our
remarks apply also to that case.

{\em Acknowledgements} I apologize to all authors whose related work
has not been mentioned here. A complete  bibliography for
the discussed topics greatly exceeds 50 papers.

{\bf Figure Captions}\\

\noindent Fig.1. Contour plot of $\delta^{e^{+}e^{-}}$ for
$s/\Lambda_{\overline{MS}}^{2}=75^{2}$, as a function of the scheme
parameters, obtained by evaluating the contour integral numerically.
The regions of allowed scheme parameters for $l=2,3,6,10$ are
also indicated.\\

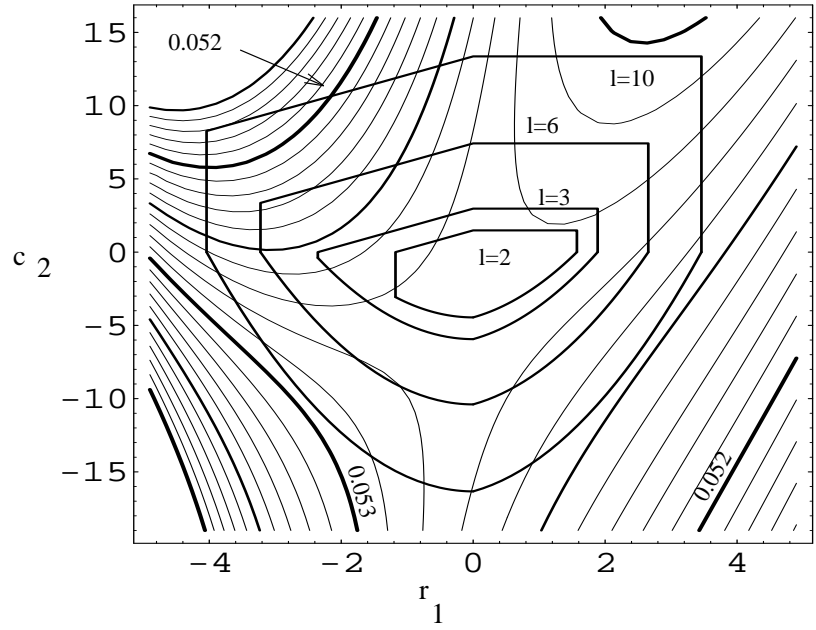
\noindent Fig.2. QCD correction to the Bjorken sum rule for
$n_{f}=3$, as a function of $\sqrt{Q^{2}}/\Lambda_{\overline{MS}}$,
for different values of $r_{1}$ and $c_{2}$ belonging to the $l=2$
allowed region: a)(-1.65,0), b)(0.5,5.27), c)(-1.0,5.27),
d)(-1.65,5.27). The dashed curve indicates prediction in the
$\overline{MS}$ scheme.\\


\begin{thebibliography}{99}

\bibitem{DukeRoberts} For a summary of early contributions see
 D. W. Duke and R. G. Roberts, Phys. Rep. {\bf 120}, 275 (1985).

\bibitem{PMS} P. M. Stevenson,  Phys. Lett. {\bf 100B}, 61 (1981),
 Phys. Rev. {\bf D23}, 2916  (1981).

\bibitem{par95} P. A. R\c{a}czka, Zeitschrift f\"ur Physik {\bf C65},
 481 (1995).

\bibitem{par92} P. A. R\c{a}czka,  Phys. Rev. {\bf D 46}
 R3699  (1992), see also P. A. R\c{a}czka, Proceedings of the XV
International Warsaw Meeting on Elementary Particle Physics,
Kazimierz, Poland, 25-29 May 1992, edited by Z. Ajduk, S. Pokorski and
A. K. Wr\'{o}blewski (World Scientific, Singapore, 1993), p.\ 496.

\bibitem{bra89} E. Braaten, Phys. Rev. {\bf D39}, 1458 (1989).

\bibitem{kataev} S. G. Gorishny, A. L. Kataev and S. A. Larin,
 Phys. Lett. {\bf B259}, 144 (1991).

\bibitem{samuel} M. A. Samuel and L. R. Surguladze, Phys. Rev.
 {\bf D44}, 1602 (1991).

\bibitem{piv} A. A. Pivovarov, Z. Phys. {\bf C53}, 461 (1992).

\bibitem{ledepi} F. LeDiberder and A. Pich, Phys. Lett. {\bf B286}, 147
 (1992).

\bibitem{parszym94} P. A. R\c{a}czka and A. Szymacha, preprint
 IFT/13/94, (hep-ph/9412236, revised).

\bibitem{radu} A. Radyushkin, JINR preprint E2-82-159 (1982).

\bibitem{parszym95} P. A. R\c{a}czka and A. Szymacha, preprint
 IFT/1/95.

\bibitem{larin} S. A. Larin and J. A. M. Vermaseren, Phys.\ Lett.\
 {\bf B259}, 345 (1991).
\end{thebibliography}
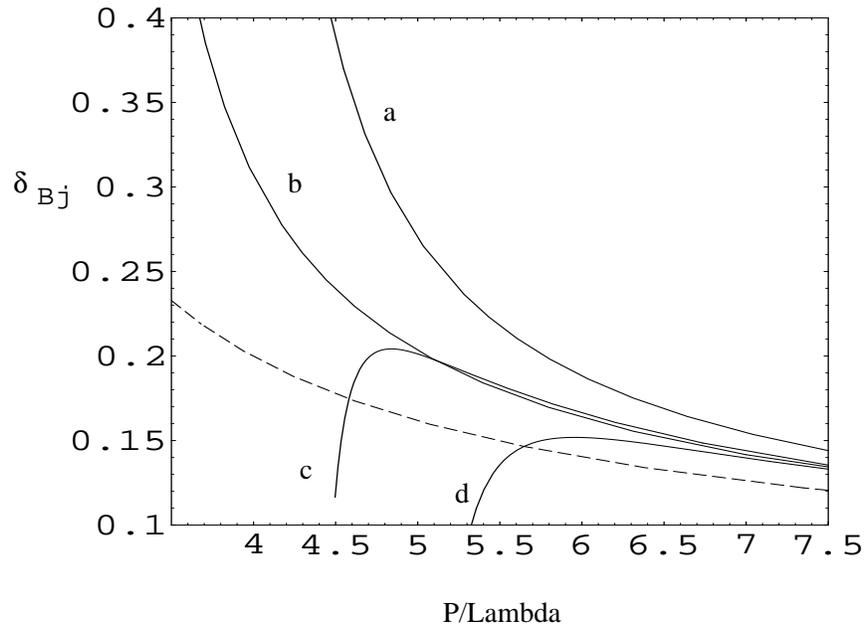
\end{document}